\documentstyle[12pt]{article}
\begin{document}
\hfill UTHEP--94--0703\vskip.01truein
\hfill{July 1994}\vskip1truein
\def\fnote#1#2{
\begingroup\def\thefootnote{#1}\footnote{#2}\addtocounter{footnote}{-1}
\endgroup}
\def\dslash{\not{\hbox{\kern-2pt $\partial$}}}
\def\eslash{\not{\hbox{\kern-2pt $\epsilon$}}}
\def\Dslash{\not{\hbox{\kern-4pt $D$}}}
\def\Aslash{\not{\hbox{\kern-4pt $A$}}}
\def\Qslash{\not{\hbox{\kern-4pt $Q$}}}
\def\pslash{\not{\hbox{\kern-2.3pt $p$}}}
\def\kslash{\not{\hbox{\kern-2.3pt $k$}}}
\def\qslash{\not{\hbox{\kern-2.3pt $q$}}}
\def\np#1{{\sl Nucl.~Phys.~\bf B#1}}\def\pl#1{{\sl Phys.~Lett.~\bf #1B}}
\def\pr#1{{\sl Phys.~Rev.~\bf D#1}}\def\prl#1{{\sl Phys.~Rev.~Lett.~\bf
#1}}
\def\cpc#1{{\sl Comp.~Phys.~Comm.~\bf #1}}
\def\anp#1{{\sl Ann.~Phys.~(NY) \bf #1}}\def\etal{{\it et al.}}
\def\half{{\textstyle{1\over2}}}
\def\be{\begin{equation}}\def\ee{\end{equation}}
\def\ba{\begin{array}}\def\ea{\end{array}}
\def\tr{{\rm tr}}
\centerline{\Large Particle Hamiltonian and Perturbative Gauge Theories}
\vskip.8truein
\centerline{\sc George Siopsis\fnote{\ast}{e-mail: {\tt
siopsis@panacea.phys.utk.edu}}}
\vskip.5truein
\centerline{\it Department of Physics and Astronomy}
\centerline{\it The University of Tennessee, Knoxville, TN 37996--1200}
\centerline{\it U. S. A.}\vskip.05truein\baselineskip=21pt\vskip.6truein
\centerline{\bf ABSTRACT}\vskip.2truein\par\let\sstl=\scriptscriptstyle
We discuss the Hamiltonian formulation of the Schwinger proper-time method of
calculating Green functions in gauge theories. Instead of calculating Feynman
diagrams, we solve the corresponding Dyson-Schwinger equations. We express the
solutions in terms of vertex operators, which consist of the zero modes of the
vertex operators appearing in string theories. We show how color decomposition
arises in this formalism at tree level. At the one-loop level, we arrive at
expressions similar to those obtained in the background gauge. In both cases,
no special gauge-fixing procedure is needed.
\par\renewcommand\thepage{}
\vfill\eject\parskip.1truein
\pagenumbering{arabic}\par


It has always been difficult to compare theory with experimental data of strong
interactions, because calculations of higher-order diagrams in non-abelian
gauge
theories are very complicated. In recent years, considerable progress has been
made in simplifying these calculations, by introducing new techniques (spinor
helicity method~\cite{shm}, color decomposition~\cite{codc}, supersymmetry
identities~\cite{suid}, recursion relations~\cite{rere}, electric circuit
analogy~\cite{ref16}, etc). An interesting approach has been developed by Bern,
{\em et al.}, and is based on string theory~\cite{ref2}. They exploited the
organization of string theory diagrams to formulate rules for the calculation
of
diagrams in non-abelian gauge theories. Although these rules are inspired by
string theory, they can be formulated in pure field theoretical terms. The
question then arises whether similar organizing principles can be derived from
field theory. This is an important question, because such principles will take
us beyond the one-loop order in perturbation theory. They may also permit the
inclusion of fermion lines in diagrams, which is not an easy task in the
present
string theoretical formulation.

A more direct approach to the re-organization of Feynman diagrams in
non-abelian
gauge theories has been taken by Strassler~\cite{ref5}. He showed that
one-particle irreducible functions in an external gauge field may be written as
quantum mechanical amplitudes of particles moving on world lines. To this end,
he employed the Schwinger proper-time formalism~\cite{schw} and expressed the
results in terms of one-dimensional path integrals. This approach seems to rely
heavily on the existence of a Lagrangian which is either quadratic in certain
fields, or can be written in this form. Thus, it appears to be limited to
one-loop diagrams.\fnote{\dagger}{Extensions to this approach have also been
studied~\cite{ref12}.}

To go beyond effective actions, we shall use the Schwinger proper-time
formalism
in order to solve Dyson-Schwinger equations perturbatively. The solutions
cannot
always be written as path integrals of particle Lagrangians with a simple
interaction term. For explicit calculations, it seems to be advantageous to use
the Hamiltonian formalism, and express the amplitudes as expectation values of
a
string of vertex operators, which are reminiscent of the old operator formalism
in string theory.

We start by considering the case of an external gauge field. In this case, the
solution to the Dyson-Schwinger equation exponentiates, which explains why it
is
possible to express it in terms of a one-dimensional path integral. We shall
discuss this exponentiation in the Hamiltonian formalism in order to generalize
it later. For definiteness, consider a scalar field $\phi$ coupled to an
external gauge field $A^\mu =A^{\mu a} T^a$, where $T^a$ are the generators of
the algebra of the gauge group. The classical equation of motion for $\phi$ is
\be (i\partial_\mu +gA_\mu (x))^2 \phi (x)=0\;. \label{cleq}\ee
In quantum field theory, the propagator $\Delta^{AB} (x,x')=\langle \phi^A
(x)\phi^B (x') \rangle$ satisfies the corresponding Dyson-Schwinger equation,
\be (i\partial_\mu +gA_\mu (x))^2\Delta (x,x') =\delta^4
(x-x')\;,\label{ena}\ee
where we have suppressed group theory indices. In order to simplify the
notation, we shall consider the abelian case first.
We may solve Eq.~(\ref{ena}) with the aid of the Schwinger proper time $\tau$
as
follows.
Define operators $x^\mu$ and $p^\mu$ satisfying the standard commutation
relations,
\be [p^\mu \;\;,\;\; x^\nu ] = i\eta^{\mu\nu}\;.\ee
Let them describe a system whose evolution is governed by the Hamiltonian
\be H=(p_\mu +gA_\mu (x) )^2 \;.\ee
If $|q\rangle$ and $|q'\rangle$ are eigenstates of the momentum operator,
$p^\mu$, the function
\be U(q,q';\tau ) = \langle q'|e^{-iH\tau}|q\rangle \;, \ee
which is a matrix element of the evolution operator, $U(\tau) = e^{-iH \tau}$,
satisfies the equation,
\be i\partial_\tau U(q,q';\tau ) =(q_\mu +gA_\mu (x))^2U(q,q';\tau )\;,
\label{u0e}\ee
where $x^\mu =- i\partial /\partial q_\mu$.
Integrating over the proper time $\tau$, and using the orthogonality of
momentum
eigenfunctions, $\langle q|q'\rangle = (2\pi )^4
\delta^4 (q+q')$ (treating both momenta as outgoing), we obtain
\be (q_\mu +gA_\mu (x))^2 \int_0^\infty d\tau U(q,q';\tau ) = -i(2\pi )^4
\delta^4 (q+q') \;.\ee
Comparison with Eq.~(\ref{ena}) shows that $i\int_0^\infty d\tau U(q,q';\tau )$
is the scalar propagator in momentum space.
So, instead of using Feynman diagrams to calculate the propagator, we may
instead calculate the evolution function, $U(q,q';\tau )$.
For certain gauge fields, this function can be calculated exactly. However,
we want to calculate $U(q,q';\tau )$ for general fields, because we are
interested in dynamical gauge bosons. We shall therefore use perturbation
theory.

By splitting the Hamiltonian as $H=H_0+gH_I$, where
\be H_0 = p_\mu p^\mu \;\;,\;\; H_I = \{p_\mu \,,\, A^\mu \} + g A_\mu A^\mu
\;,\ee
we can write the evolution operator in terms of time-ordered products as
\be U(\tau) = \sum_{n=0}^\infty {(-ig)^n \over n!}\int_0^\tau
dt_1 \cdots \int_0^\tau dt_n T(H_I(t_1)\cdots H_I(t_n)) = T e^{-ig\int_0^\tau
dt
H_I (t)} \;,\ee
where time evolution is now governed by $H_0$.
Thus, $x^\mu (t)$ and $p^\mu (t)$ satisfy the Ehrenfest relations
\be {dx^\mu\over dt} = [H_0\,,\, x^\mu]=2ip^\mu \;\;,\;\; {dp^\mu\over dt} =
[H_0\,,\, p^\mu]=0 \;.\ee
The solution to these equations is
\be x^\mu (t) =2i\bar p^\mu t+ \bar x^\mu \;\;,\;\; p^\mu (t) = \bar p^\mu
\;,\ee
where $\bar x^\mu$ and $\bar p^\mu$ are the position and momentum operators,
respectively, at time $t=0$.
We can therefore write
\be U(q,q';\tau ) = \langle q';0 | Te^{i\int_0^\tau dt H_I (t)} |q;\tau \rangle
\;, \label{duo}\ee
where the state $|q;t\rangle$ satisfies the Schr\"odinger equation,
$i\partial_t |q;t\rangle = H_0|q;t\rangle$. It is clear from Eq.~(\ref{duo})
that $U(q,q';\tau )$ can be expressed as a quantum mechanical path integral,
with appropriate boundary conditions at times $t=0$ and $t=\tau$. This is
possible because the terms involving the interaction Hamiltonian, $H_I$, have
neatly exponentiated. We wish to generalize this to the case where such an
exponentiation is not possible, so we shall use the operator formalism
to calculate the various matrix elements in Eq.~(\ref{duo}).
By expressing $x^\mu (t)$ in terms of $x^\mu (\tau )$ and $x^\mu (0)$,
\be x^\mu (t) ={\textstyle{t\over\tau}} x^\mu (\tau ) +{\textstyle
(1-{t\over\tau})} x^\mu (0) \;,\ee
we can write $H_I$ in terms of operators $V_k (0)$ and $V_k (\tau )$, where the
vertex operator $V_k (t)$ is defined by
\be V_k (t) = e^{ik\cdot x(t)} \;. \label{vkt}\ee
This operator consists of the zero modes of the tachyonic vertex operator in
string theory.
Matrix elements containing insertions of this operator can be calculated by
using the commutation relations
\be V_{k_1} (\tau ) V_{k_2} (0) = e^{-k_1 \cdot k_2 \tau} V_{k_2} (0) V_{k_1}
(\tau ) \quad ,\quad [V_{k_1} (0) \; ,\; V_{k_2} (0)] = [V_{k_1} (\tau ) \;,\;
V_{k_2} (\tau )] =0 \;, \ee
and the properties
\be \langle q; 0| V_k (0) = \langle q+k ;0| \quad ,\quad V_k (\tau ) |q;\tau
\rangle = |q+k;\tau \rangle \;. \ee
We have now assembled all the tools which are necessary for the calculation of
$U(q,q';\tau )$. By expanding in the coupling constant,
\be U(q,q';\tau ) = U^{(0)}(q,q';\tau ) +g\,U^{(1)}(q,q';\tau ) + \dots \;,\ee
we obtain
\be U^{(0)}(q,q';\tau ) = \langle q';0 |q;\tau \rangle = e^{-\tau q^2} (2\pi
)^4
\delta^4 (q+q') \;, \label{u0}\ee
whose integral over $\tau$ is the propagator for a scalar, $1/q^2$.
To first order in the coupling constant, we have
\be U^{(1)}(q,q';\tau ) = \int_0^\tau dt \langle q';0 |
\{ p_\mu \;,\; A^\mu (t)\}|q;\tau \rangle \;. \label{u1}\ee
The projection of this functional onto a plane wave,
$A^\mu = \epsilon^\mu e^{ik\cdot x} =\epsilon^\mu V_k$,
can be written in the form
\be U_\mu^{(1)}(q,q',k;\tau ) = (q-q')_\mu \int_0^\tau dt \langle q';0 | V_k
(t)
|q;\tau \rangle \;.\ee
After some straightforward algebra, we obtain
\be U_\mu^{(1)}(q,q',k;\tau ) = (2\pi )^4 \delta^4 (q+q'+k) (q-q')_\mu
\int_0^\tau dt e^{-\tau q^2 -(2q+k) \cdot kt} \;.\ee
Upon integrating over $t$ and $\tau$, we recover the standard expression,
${1\over q^2} (q-q')_\mu {1\over q'^2}$. This can easily be obtained by
amputating the two scalar legs (which amounts to multiplying by $q^2 q'^2$),
and
integrating by parts twice.

To second order in the coupling constant, we have
\be U^{(2)}(q,q';\tau ) = \langle q';0 |\left(\int_0^\tau dt_1 \int_0^{t_1}
dt_2
\{ p_\mu \;,\; A^\mu (t_2)\} \{ p_\nu \;,\; A^\nu (t_1)\} +\int_0^\tau dt A_\mu
(t) A^\mu (t)\right) |q;\tau \rangle \;. \label{u2}\ee
Projecting onto plane waves, we obtain the evolution function
\be \ba{ll} U_{\mu\nu}^{(2)}(q,q',k_1,k_2;\tau ) = & \langle q';0 |
\Big(\int_0^\tau dt_1 \int_0^{t_1} dt_2
\{ p_\mu , V_{k_2} (t_2)\}\{ p_\nu , V_{k_1} (t_1)\}
+(\mu\leftrightarrow \nu )\cr\cr & \phantom{\langle q;\tau |} + \eta_{\mu\nu}
\int_0^\tau dt V_{k_2} (t) V_{k_1} (t)\Big) |q; \tau \rangle \;. \ea
\label{u2s}
\ee
After some straightforward manipulations, we can bring this into the form
\be \ba{ll} U_{\mu\nu}^{(2)}(q,q',k_1,k_2;\tau ) = & (2q+k_1)_{\{\mu\,,}
(2q'+k_2)_{\nu\}} \int_0^\tau dt_1 \int_0^{t_1} dt_2 e^{-\tau q^2 -t_1k_1\cdot
(2q+k_1)+t_2 k_2\cdot (2q'+k_2)}\cr\cr
&-\eta_{\mu\nu} \int_0^\tau dt e^{-\tau q^2 -tk_1\cdot (2q+k_1) +tk_2\cdot
(2q'+k_2)} \;, \ea\ee
which can be shown to agree with the expression obtained from Feynman diagrams.
To obtain the one-loop scalar contribution to the self-energy of gauge bosons,
we need to sew the two scalar legs together, after amputating one of them. We
therefore need to calculate the quantity
\be U_{\mu\nu}^{(2)} (k_1,k_2; \tau ) \equiv \int {d^4q \over (2\pi )^4} \int
{d^4 q'\over (2\pi )^4} q^2 (2\pi )^4 \delta^4 (q+q') \tr U_{\mu\nu}^{(2)}
(q,q',k_1,k_2;\tau ) \;. \ee
To calculate $U_{\mu\nu}^{(2)} (k_1,k_2; \tau )$, first we integrate by parts
(using $q^2 e^{-\tau q^2} = -{d\over d\tau} e^{-\tau q^2}$), to obtain
\be \ba{ll} U_{\mu\nu}^{(2)}(k_1,k_2;\tau ) = & -\int {d^4q \over (2\pi )^4}
(2q+k_1)_{\{\mu\,,} (2q+k_2)_{\nu\} } \int_0^\tau dt e^{-\tau q^2 -(\tau -t)
k_1\cdot (2q+k_1)}\cr\cr
&+\eta_{\mu\nu} \int {d^4q \over (2\pi )^4} e^{-\tau q^2} \;, \ea\ee
where we discarded a total derivative with respect to $\tau$, and omitted a
$\delta$-function imposing conservation of momentum ($(2\pi )^4 \delta^4
(k_1+k_2)$). It is now easy to integrate over the momentum by completing the
square in the exponent. The result is
\be U_{\mu\nu}^{(2)} (k_1,k_2; \tau ) = -{1\over 16\pi^2} \left\{ \int_0^\tau
dt
\left( {\textstyle{1\over\tau^2}} \eta_{\mu\nu} + {\textstyle{1\over 2\tau}}
(1-2{\textstyle{t\over \tau}} )^2 k_{1\mu} k_{1\nu} \right) e^{-k_1^2
t(1-t/\tau
)} - {\textstyle{1\over\tau}}\eta_{\mu\nu} \right\} \;. \label{u2f}\ee
By integrating the first term by parts, we can write $U_{\mu\nu}^{(2)}
(k_1,k_2;
\tau )$ in a manifestly gauge invariant form,
\be U_{\mu\nu}^{(2)} (k_1,k_2; \tau ) = - {1\over 32 \pi^2 \tau} (k_{1\mu}
k_{1\nu} - \eta_{\mu\nu} k_1^2) \int_0^\tau dt (1-2{\textstyle{t\over \tau}}
)^2
e^{-k_1^2 t(1-t/\tau )} \;. \label{u2gi}\ee
Next, we consider the case of a fermion, $\psi$, for which the propagator in
momentum space,
$\Delta (q,q')$ satisfies the Dyson-Schwinger equation
\be (\qslash + g \Aslash (x)) \Delta (q,q') = (2\pi )^4 \delta^4 (q+q') \;,\ee
where $x^\mu = -i\partial /\partial q_\mu$.
We can express $\Delta (q,q')$ in terms of $\Delta'(q,q')$, where
\be \Delta (q,q')= \qslash \Delta' (q,q') \;.\label{deltp}\ee
Then $\Delta'(q,q')$ satisfies the equation
\be \left(q^2 + g \Aslash (x)\qslash \right)
\Delta' (q,q') = (2\pi )^4 \delta^4 (q+q') \;. \ee
To calculate $\Delta' (q,q')$, we may follow the same procedure as before,
except that
now the interaction Hamiltonian is simpler,
\be H_I = \Aslash (x) \pslash \;.\ee
Notice that we did not have to use the second-order formalism for fermions,
which would amount to introducing the propagator $\Delta'' (q,q')$, defined by
\be \Delta (q,q')= (\qslash + g \Aslash (x)) \Delta'' (q,q') \;, \ee
instead of $\Delta'(q,q')$ (Eq.~(\ref{deltp})). The latter leads to more
elegant
expressions, but the interaction Hamiltonian is more complicated. As a result,
our computations will be simpler than those in the path integral approach for
fermions~\cite{ref5}. Of course, the final expressions in the two approaches
agree with each other.

The $n$th-order contribution to the evolution function $U(q,q'; \tau )$ can be
written as a time-ordered matrix element, as before,
\be U^{(n)}(q,q';\tau ) = \int_0^\tau dt_1 \cdots \int_0^{t_{n-1}} dt_n \langle
q';0 | \Aslash (t_n) \pslash \cdots \Aslash (t_1) \pslash |q;\tau \rangle \;.
\ee
To give an example of an explicit calculation, consider
the one-loop fermion contribution to the self-energy of gauge bosons,
\be U_{\mu\nu}^{\rm 1-loop} (k_1, k_2 ;\tau ) = \int {d^4q \over (2\pi )^4}
\int
{d^4q' \over (2\pi )^4} q^2 (2\pi )^4 \delta^4 (q+q') \tr U_{\mu\nu}^{(2)} (q,
q', k_1, k_2; \tau ) \;, \ee
where ({\em cf.}~Eq.(\ref{u2s}))
\be U_{\mu\nu}^{(2)} (q, q', k_1, k_2; \tau ) = \int_0^\tau dt_1 \int_0^{t_1}
dt_2 \langle q'; 0| \left( \gamma_\mu V_{k_2} (t_2) \pslash \gamma_\nu V_{k_1}
(t_1) \pslash + (\mu \leftrightarrow \nu ) \right) |q; \tau \rangle \;. \ee
Manipulating this expression as in the scalar case, we obtain
\be U_{\mu\nu}^{(2)} (k_1,k_2; \tau ) = {1\over 4\pi^2} (k_{1\mu} k_{1\nu} -
\eta_{\mu\nu} k_1^2 ) \int_0^\tau dt {\textstyle ( 1- {t\over\tau} )
{t\over\tau}} e^{-k_1^2 t(1-t/\tau )} (2\pi )^4 \delta^4 (k_1+k_2) \;,\ee
which is in a manifestly gauge-invariant form without any further
manipulations.

Next, we turn to the calculation of amplitudes involving dynamical gauge
bosons.
In order to calculate amplitudes, we need to fix the gauge. We shall choose the
Lorentz-Feynman gauge, in which
\be \partial^\mu A_\mu^a =0 \;, \ee
and the propagator for the gauge field is
\be \langle A_\mu^a (k) A_\nu^b (k') \rangle = i\delta^{ab} \eta_{\mu\nu}
{1\over k^2} (2\pi )^4 \delta^4 (k-k') \;.\label{gpro} \ee
First, consider the gauge coupling of a scalar field. Let
\be \Gamma_\mu^{ABa} (q,q', k) = \langle \phi^A (q) \phi^B (q') A_\mu^a (k)
\rangle \label{thrp}\ee
be the three-point function, with the two scalar legs attached. To lowest order
in the coupling constant, it satisfies the Dyson-Schwinger equation (in matrix
notation)
\be q^2 \Gamma_\mu^a (q,q'; k) = g T^a \{ q_\mu \;,\; e^{ik\cdot x} \} \Delta
(q,q') \;, \label{ug0}\ee
where $x^\mu = -i\partial /\partial q_\mu$ and $\Delta (q,q')$ is the scalar
propagator, corresponding to the evolution function $U^{(0)} (q,q'; \tau )$
(Eq.~(\ref{u0})). Eq.~(\ref{ug0}) may be solved by introducing the Schwinger
proper time. To use previous results, notice that $U_\mu^{(1)} (q,q',k; \tau )$
(Eq.~(\ref{u1})) obeys the equation
\be i\partial_\tau U_\mu^{(1)a} (q,q',k; \tau ) = q^2 U_\mu^{(1)a} (q,q',k;
\tau
) + T^a \{ q_\mu \;,\; e^{ik\cdot x} \} U^{(0)} (q,q'; \tau ) \;,\ee
which is a consequence of Eq.~(\ref{u0e}). It follows that
\be \Gamma_\mu (q,q'; k) = ig\int_0^\infty d\tau U_\mu^{(1)} (q,q',k; \tau )
\;.
\ee
Therefore, the three-point correlation function corresponds to the first-order
evolution function, $U^{(1)} (q,q';\tau )$, even if the gauge boson is a
dynamical field. This simple correspondence extends to higher orders in the
abelian case (as may be seen recursively, by using the appropriate
Dyson-Schwinger equation for $\langle \phi (q) \phi (q') A^{\mu_1} \cdots
A^{\mu_n} \rangle$), but fails for non-abelian groups, because of the gauge
boson self-coupling. Yet, higher-order non-abelian diagrams can still be
described by proper-time integrals. We shall demonstrate this in the case of a
pure Yang-Mills theory. It will then be clear how one can extend the formalism
to include matter fields.

Classically, the gauge field satisfies the non-abelian Maxwell equation, which
is conveniently written in matrix form in the Lorentz-Feynman gauge,
\be \Box A_\mu = 2g [A_\sigma \,,\, \partial^\sigma A_\mu ] +g [\partial^\sigma
A_\sigma \,,\, A_\mu ] -g [A^\sigma \,,\, \partial_\mu A_\sigma ] + g^2 [
A^\sigma \,,\, [ A_\sigma \,,\, A_\mu ]] \label{maxe}\ee
To write down the Dyson-Schwinger equation for the three-point coupling,
\be G_{\mu\nu\rho}^{abc} (k_1,k_2,k_3) = \langle A_\mu^a (k_1) A_\nu^b (k_2)
A_\rho^c (k_3) \rangle \;, \ee
we need to attach the leg with momentum $k_1$. The function
\be G_{\mu\nu\rho}^{\ast bc} (k_1,k_2,k_3) = {\eta_\mu^{\phantom\mu\sigma}
\over
k_1^2} T^a G_{\sigma\nu\rho}^{abc} (k_1,k_2,k_3) \ee
satisfies the equation
\be k_1^2 G_{\mu\nu\rho}^{\ast bc} (k_1,k_2,k_3) = g \, T^b T^c k_3^2
(\eta_{\mu\nu} (k_1-k_2)_\rho + \eta_{\nu\rho} (2k_2+k_1)_\mu - \eta_{\rho\mu}
(2k_1+k_2)_\nu ) e^{ik_2\cdot x} \Delta ( k_1,k_3) + \dots \;,\ee
where $x_\mu =-i\partial /\partial k_1^\mu$, and $\delta^{ab} \eta_{\mu\nu}
\Delta$ is the propagator for the gauge field (Eq.~(\ref{gpro})). The dots
denote a term which is obtained from the first one by a permutation of the two
amputated legs (color decomposition~\cite{codc}). It can easily be seen that
the
solution to this equation can be written in the form
\be G_{\mu\nu\rho}^{\ast bc} (k_1,k_2,k_3) = k_3^2 {\cal A}_{\mu\nu\rho}^{\ast
bc} (k_1,k_2,k_3) + k_2^2 {\cal A}_{\mu\rho\nu}^{\ast cb} (k_1, k_3, k_2) \;,
\ee
where ${\cal A}^\ast$ is the color-ordered partial amplitude,
\be {\cal A}_{\mu\nu\rho}^{\ast bc} (k_1,k_2,k_3) = ig \, T^b T^c k_3^2
\int_0^\infty d\tau \int_0^\tau dt \langle k_1;0| {\cal H}_{\mu\nu\rho}^{(1)}
(k_2; t) |k_3;\tau \rangle \;, \label{pram} \ee
and
\be {\cal H}_{\mu\nu\rho}^{(1)} (k ; t ) = \eta_{\mu\nu} (2p_\rho V_k (t) -V_k
(t) p_\rho) + \eta_{\nu\rho} (2V_k (t) p_\mu - p_\mu V_k (t)) +\eta_{\rho\mu}
\{
V_k (t) \,,\, p_\nu \} \;. \label{h1na}\ee
Manipulating this expression as before, we recover the standard expression,
\be G_{\mu\nu\rho}^{abc} (k_1,k_2,k_3) = ig \, f^{abc} (\eta_{\mu\nu}
(k_1-k_2)_\rho + \eta_{\nu\rho} (k_2-k_3)_\mu + \eta_{\rho\mu} (k_3-k_1)_\nu )
+
\dots \;. \ee
Similarly, the four-point amplitude with one leg attached satisfies the
equation
\be \ba{ll} k_1^2 G_{\mu\nu\rho\lambda}^{\ast bcd} (k_1, \dots ,k_4) = & ig
\int
d^4 x  e^{ik_1\cdot x} \,\{ \, [ T^{b'}\,,\, T^{c'} ] \langle  (2A_\sigma^{b'}
(x) \partial^\sigma A_\mu^{c'} (x) + \partial^\sigma A_\sigma^{b'} (x)
A_\mu^{c'} (x) \cr\cr & \quad\quad - A^{b'\sigma} (x) \partial_\mu
A_\sigma^{c'}
(x)) A_\nu^b (k_2) A_\rho^c (k_3) A_\lambda^d (k_4) \rangle \cr\cr & \quad\quad
+g [T^{b'} \,,\, [T^{c'} \,,\, T^{d'} ]] \langle A^{b'\sigma} (x) A_\sigma^{c'}
(x) A_\mu^{d'} (x) A_\nu^b (k_2) A_\rho^c (k_3) A_\lambda^d (k_4) \rangle \,\}
\;. \ea \ee
The various terms are conveniently organized by color ordering. The result is
\be \ba{ll} k_1^2 G_{\mu\nu\rho\lambda}^{\ast bcd} (k_1,\dots ,k_4) & = ig T^b
(\eta_{\mu\nu} (k_1-k_2)_\sigma + \eta_{\nu\sigma} (2k_2+k_1)_\mu -
\eta_{\sigma\mu} (2k_1+k_2)_\nu ) e^{ik_2\cdot x} {\cal
A}_{\sigma\rho\lambda}^{\ast cd} (k_1, k_3,k_4) \cr\cr &+ ig e^{ik_3\cdot x}
{\cal A}_{\mu\nu\sigma}^{\ast bc} (k_1,k_2,k_4) T^d (\eta_{\sigma\rho}
(k_1-k_3)_\lambda + \eta_{\rho\lambda} (2k_3+k_1)_\sigma - \eta_{\lambda\sigma}
(2k_1+k_3)_\rho )\cr\cr &+g^2 T^b T^c T^d (\eta_{\nu\rho} \eta_{\mu\lambda} +
\eta_{\mu\rho} \eta_{\lambda\nu} -2\eta_{\lambda\rho} \eta_{\mu\nu}) e^{ik_2
\cdot x} e^{ik_3\cdot x} \Delta (k_1, k_4) \cr\cr &+ \dots \;,\ea
\label{g4a}\ee
where $x_\mu = -i\partial /\partial k_1^\mu$.
It is interesting to note that the last term on the right-hand side corresponds
to the Feynman rule for the four-point gauge coupling obtained by string
theoretical techniques~\cite{ref2}. This is obtained in our formalism without
any special gauge fixing procedure. The solution to Eq.~(\ref{g4a}) can be
written in terms of partial amplitudes ({\em cf.}~Eq.~(\ref{pram})) which are
integrals over proper time $\tau$, of
\be \ba{ll} U_{\mu\nu\rho\lambda}^{(2)} (k_1, \dots ,k_4) &= \langle k_1 ; 0|
\Big( \int_0^\tau dt_1 \int_0^{t_1} dt_2 {\cal H}_{\mu \nu \sigma}^{(1)}
(k_2;t_2) {\cal H}_{\phantom\sigma\rho\lambda}^{\sigma (1)} (k_3;t_1) \cr\cr &
\phantom{= \langle k_1 ; 0|} + \int_0^\tau dt {\cal
H}_{\mu\nu\rho\lambda}^{(2)}
(k_2,k_3) \Big) |k_4 ;\tau\rangle + \dots \;, \ea \label{u2h}\ee
multiplied by the appropriate generators of the gauge group,
where ${\cal H}_{\mu\nu\rho}^{(1)} (k;t)$ is given by Eq.~(\ref{h1na}), and
\be {\cal H}_{\mu\nu\rho\lambda}^{(2)} (k_1,k_2) = (\eta_{\nu\rho}
\eta_{\mu\lambda} + \eta_{\mu\rho} \eta_{\lambda\nu} -2\eta_{\lambda\rho}
\eta_{\mu\nu}) V_{k_1} (t) V_{k_2} (t) \;. \ee
Clearly, Eq.~(\ref{u2h}) does not correspond to a quantum-mechanical
interaction
Hamiltonian. Yet, the four-point amplitude can be written in terms of the
vertex
operator (Eq.~(\ref{vkt})).

Loop diagrams can also be calculated in the same fashion.
The one-loop contribution to the gauge boson self-energy in a pure Yang-Mills
theory, $\Gamma_{\mu\nu}^{ab} (k_1,k_2)$, may be calculated by using the
Dyson-Schwinger equation
\be \ba{ll}  k_1^2 \Gamma_{\mu\nu}^{\ast b} (k_1,k_2) =& ig\int d^4 x\;
e^{ik_1\cdot x} \, \Big\{ \, [T^{b'} \,,\, T^{c'} ] \langle (2A_\sigma^{b'}
\partial^\sigma A_\mu^{c'} + \partial^\sigma A_\sigma^{b'} A_\mu^{c'} -
A^{b'\sigma} \partial_\mu A_\sigma^{c'}) (x) A_\nu^{b'} (k_2) \rangle \cr\cr &
\quad\quad g [T^{b'} \,,\, [ T^{c'} \,,\, T^{d'} ]] \langle A^{b'\sigma} (x)
A_\sigma^{c'} (x) A_\mu^{d'} (x) A_\nu^b (k_2) \rangle \cr\cr & \quad\quad +
[T^{b'} \,,\, T^{c'} ] \langle \overline \omega^{b'} (x) i\partial_\mu
\omega^{c'} (x) A_\nu^b (k_2) \rangle \,\Big\} \;, \label{onl}\ea\ee
where the last term accounts for the effects of the ghosts. It can be
calculated
using the appropriate Dyson-Schwinger equation. Eq.~(\ref{onl}) may
be written as
\be \ba{ll}  k_1^2 \Gamma_{\mu\nu}^{\ast b} (k_1,k_2) &= ig\int \widetilde{dq}
\Big\{ (q_\mu \eta^{\rho\sigma} -(q+k_1)^\sigma \eta_\mu^{\phantom\mu\rho} )
e^{ik_1\cdot x} [ G_{\sigma\rho\nu}^{\ast c'b}(q,q',k_2) \;,\; T^{c'} ] \cr\cr
&
\phantom{= ig\int \widetilde{dq}}+ g [T^{b'} \,,\, [ T^{b'} \,,\, T^b ]]
e^{ik_1\cdot x} e^{ik_2 \cdot x} \eta_{\mu\nu} \Delta (q,q') \cr\cr
& \phantom{= ig\int \widetilde{dq}} +e^{ik\cdot x} q_\mu G_\nu^{{\rm (gh)} b}
(q,q',k_2) \Big\} \;, \ea \ee
where $\widetilde{dq} = {d^4 q \over (2\pi )^4} {d^4 q'\over (2\pi )^4} (2\pi
)^4 \delta^4 (q+q')$, and $x_\mu =-i \partial /\partial q^\mu$. The solution is
\be \Gamma_{\mu\nu}^{a b} (k_1,k_2) = g^2 f^{ab'c'} f^{bb'c'} \int_0^\infty
d\tau  U_{\mu\nu}^{\rm 1-loop} (k_1,k_2; \tau ) \;, \ee
where
\be U_{\mu\nu}^{\rm 1-loop} (k_1,k_2; \tau ) = \int {d^4 q \over (2\pi )^4} q^2
\langle q;0| {\cal H}_{\mu\nu}^{\rm 1-loop} (k_1,k_2;\tau ) |q;\tau \rangle
\;,\ee
and
\be \ba{ll} {\cal H}_{\mu\nu}^{\rm 1-loop} (k_1,k_2;\tau ) &= \int_0^\tau dt_1
\int_0^{t_1} dt_2 \Big( {\cal H}_{\mu\rho\sigma}^{(1)} (k_2;t_2) {\cal
H}_{\phantom {\rho\sigma}\nu}^{\rho\sigma (1)} (k_1;t_1) + {\cal H}_\mu^{\rm
(gh)} (k_2;t_2) {\cal H}_\nu^{\rm (gh)} (k_1;t_1) \Big) \cr\cr
& + \int_0^\tau dt \eta_{\mu\nu} V_{k_2} (t) V_{k_1} (t) \;, \ea \ee
where ${\cal H}^{(1)}$ is given by Eq.~(\ref{h1na}), and
\be {\cal H}_\mu^{\rm (gh)} (k;t) = p_\mu V_k (t) \ee
is the interaction Hamiltonian corresponding to the ghosts.
After performing the operator algebra and integrating over the momentum as in
the scalar case, we obtain ({\em cf.}~Eq.~(\ref{u2f}))
\be \ba{ll} \Gamma_{\mu\nu}^{a b} (k_1,k_2) &= {g^2 \over 16\pi^2} f^{ab'c'}
f^{bb'c'} \int_0^\infty d\tau \Big\{ \int_0^\tau dt \Big( {1\over\tau^2}
\eta_{\mu\nu} + {1\over 2\tau} (1-2{t\over \tau} )^2 k_{1\mu} k_{1\nu} \cr\cr
& - {2\over \tau} (k_{1\mu} k_{1\nu} -\eta_{\mu\nu} k_1^2) \Big) e^{-k_1^2
t(1-t/\tau )} - {1\over\tau}\eta_{\mu\nu}\Big\} (2\pi )^4 (k_1+k_2) \;, \ea
\label{g2na}\ee
which can be cast into a manifestly gauge-invariant form by integrating by
parts
({\em cf.}~Eq.~(\ref{u2gi})). This expression is similar to the one obtained in
the path-integral formalism, if the background gauge is used~\cite{ref5}.
However, we arrived at Eq.~(\ref{g2na}) without any special gauge-fixing
procedure.

Matter fields can also be included in this formalism. The Maxwell equation
(\ref{maxe}) is modified by the addition of a current, which for a scalar field
is
\be j_\mu^a = \phi^\ast T^a (\partial_\mu -ig A_\mu ) \phi - {\rm c.~c.} \;.
\ee
This term introduces a correction to the Dyson-Schwinger equation for the gauge
boson self energy (Eq.~(\ref{onl})). It can be calculated similarly to the
other
terms. It is also possible to derive expressions for graphs with external
matter
legs, by using the Dyson-Schwinger equation corresponding to the Maxwell
equation, if the graph contains external gauge bosons, or the classical
equation
of motion for the matter field (Eq.~(\ref{cleq})). We have demonstrated this by
calculating the three-point coupling (Eq.~(\ref{thrp})). The calculation of
higher-order amplitudes can be done by a recursive use of Dyson-Schwinger
equations.

In conclusion, we presented a method for the calculation of Green functions in
non-abelian gauge theories, which is based on the Schwinger proper time
formalism~\cite{schw}, rather than Feynman diagrams. Using Dyson-Schwinger
equations, we arrived at expressions which resemble those derived by string
theoretical techniques~\cite{ref2}. Similar expressions for one-particle
irreducible functions have also been derived by expressing the effective action
as a path integral for a single particle moving on a world-line~\cite{ref5}.
The
method presented here is more generally applicable than the one based on the
path integral. It can be applied to tree diagrams as well as loops.
Higher-order
diagrams can be expressed in terms of lower-order ones. Thus recursively, any
diagram can be written in terms of vertex operators which consist of the zero
modes of the vertex operators in string theory. At tree level, color
decomposition arises without any special gauge-fixing procedure. At the
one-loop
level, we obtained expressions similar to those obtained in the background
gauge. It is hoped that this formalism will lead to more compact and manageable
expressions for higher-order corrections to Green functions.




\end{document}